\documentstyle[11pt,newpasp,twoside,psfig]{article}
\reversemarginpar

\begin{document}
\title{A Search for OH Megamasers at $z>0.1$:  Preliminary Results}
\author{Jeremy Darling \& Riccardo Giovanelli}
\affil{Cornell University, Department of Astronomy, Ithaca, NY  14853}

\begin{abstract}
We report the discovery of 11 OH megamasers and one OH absorber, along
with upper limits on the OH luminosity of 53 other luminous infrared
galaxies at $z > 0.1$.  The new megamasers show a wide range of spectral
properties, but are consistent with the extant set of 55 previously
reported objects, 8 of which have $z > 0.1$.  The
new OH detections are the preliminary results of a OH megamaser survey
in progress at the Arecibo Observatory\footnote{The Arecibo
Observatory is part of the National Astronomy and Ionosphere Center, which 
is operated by Cornell University under a cooperative agreement with the
National Science Foundation.}, which is expected
to produce several dozen detections and calibrate the 
luminosity function of OH megamasers to the low-redshift
galaxy merger rate ($0.1 < z < 0.2$).
\end{abstract}

\section{Introduction}

OH megamasers (OHMs) are found in luminous infrared galaxies, and strongly
favor ultraluminous infrared galaxies (ULIRGs).  
Since photometric surveys have demonstrated that nearly all ULIRGs are 
the products of major galaxy mergers (Sanders, Surace, \& Ishida 1999), 
OHMs offer a promising means to detect merging galaxies across the greater
part of the history of the universe ($0 < z < 3$).  The galaxy merger rate
plays a central role in the process of galaxy evolution, and must be 
empirically determined in order disentangle galaxy number evolution from 
luminosity evolution (Le F\'{e}vre et al. 1999).  

There are a number of
observable merger products which surveys can use to measure the galaxy
merger rate, but each has its limitations.  The most successful methods
used to date are optical surveys for the morphological 
signatures of mergers such as tidal tails, rings, shells, and filaments.
Le F\'{e}vre et al. (1999) has used this method as well as close pair
surveys to determine the merger fraction of galaxies
up to $z=0.91$.  Optical surveys, however, require high angular resolution,
suffer from dust extinction effects, are biased by the optical brightening
which occurs during major mergers, and may exclude advanced mergers.  
Other survey methods capitalize on 
the very effects which bias optical surveys.  FIR surveys for ULIRGs can
use infrared brightening to identify mergers, and this technique can
be extended down to the sub-mm regime where several groups have detected
sources which may be the high-redshift counterparts of local ULIRGs 
(Smail et al. 1999).  Molecular emission lines are also
enhanced by mergers, and some of these lines can mase, such as H$_2$O or 
OH hyperfine transitions.  OHMs, in particular, appear to be quite common 
in ULIRGs: roughly 1 in 6 ULIRGs
are observed to host OHMs (Darling \& Giovanelli 1999) --- 
it may be that {\it all} ULIRGs host OHMs, 
but the beaming of OHM emission determines the observable fraction.
Although the most distant OHM detected to date has $z=0.265$ 
(Baan et al. 1992),  the most luminous 
OHMs should be detectable up to $z \sim 3$ with current instruments 
(Briggs 1998).  
Understanding the relationship between the OHM and ULIRG luminosity
functions at low redshift should allow one to measure, via OHM surveys, 
the merger rate of galaxies across a significant
fraction of the epoch of galaxy evolution.  One goal of this
survey is to carefully quantify the OHM fraction in ULIRGs at 
$0.1 < z < 0.2$.

\section{Survey Sample and Preliminary Results}

OHM candidates were selected from the {\it IRAS} Point Source Catalog
redshift survey (PSCz; W. Saunders 1999, private communication) following
the criteria:  (1) {\it IRAS} 60 $\mu$m detection, (2) $0.1 < z < 0.45$,
and (3) $0^\circ < \delta < 37^\circ$ (the Arecibo sky).  The lower limit
on redshift is set to avoid local radio frequency interference (RFI), and
the upper limit is set by the bandpass of the L-band receiver at 
Arecibo.  These candidate selection criteria limit the number of candidates
in the PSCz to 296.  The lower bound on redshift and the requirement of
{\it IRAS} detection at 60 $\mu$m select candidates which are (U)LIRGs.

From a set of 69 candidate OHMs which were observed at Arecibo, 
11 new OHMs and 1 new OH absorber have been detected.  We can place
upper limits on the OH emission from 53 of the non-detections, while
the remaining 4 candidates remain ambiguous due to strong RFI or 
a strong radio continuum which produced standing waves in the 
bandpass and confounded line detection.  Spectra and observed and 
derived parameters of the  
new detections are reported in Darling \& Giovanelli (1999).
The new OHMs span a wide range of spectral
shapes, luminosities, masing conditions, and host properties.  The diversity 
of the OHM spectra represent a corresponding diversity of physical
conditions in the masing regions, possibly ranging from spatially extended
($> 100$ pc) unsaturated emission associated with starbursts 
to small-scale ($< 1$ pc) saturated emission associated with AGN.

Figure 1 depicts the observed sample in a FIR color-luminosity
plot.  The distribution indicates that OHMs strongly favor the most 
FIR-luminous hosts, and have a lesser tendency to also favor ``warmer'' 
hosts.  Selection
effects influence these trends, particularly Malmquist bias.  The 
non-detections in the lowest $L_{FIR}$ range tend to have the least 
confident OH non-detection thresholds.  We estimate from the Malmquist
bias-corrected $L_{FIR}$-$L_{OH}$ relation calculated by Kandalian (1996)
that there are perhaps 4 additional OHMs lurking among the non-detections
of this sample.  This estimate is highly uncertain due to the statistics
of small numbers, the uncertainty in the $L_{FIR}$-$L_{OH}$ relation, and
the scatter of the known OHMs about the relation.  

\section{Expectations and Conclusions}

Survey work performed to date shows an OHM detection rate of 1 in 6, which 
indicates that we can expect to discover about 50 new OHMs in the 
complete survey.  This figure doubles the OHM sample and increases
the $z>0.1$ sample by a factor of six.  Based on the set of previously
reported OHMs, we can also expect to discover a few OH ``gigamasers''
($L_{OH} > 10^4 L_\odot$).  There are two of these extremely luminous
OHMs known today.   Detection of more OH gigamasers will solidify
confidence in upper end of the OHM luminosity function and will lend 
merit to the notion that OH gigamasers should be observable up to $z\sim3$ 
with current instruments (Briggs 1998).  The existence of OH gigamasers
is crucial to the study of the galaxy merger rate at high redshifts.
\begin{figure}
\plotfiddle{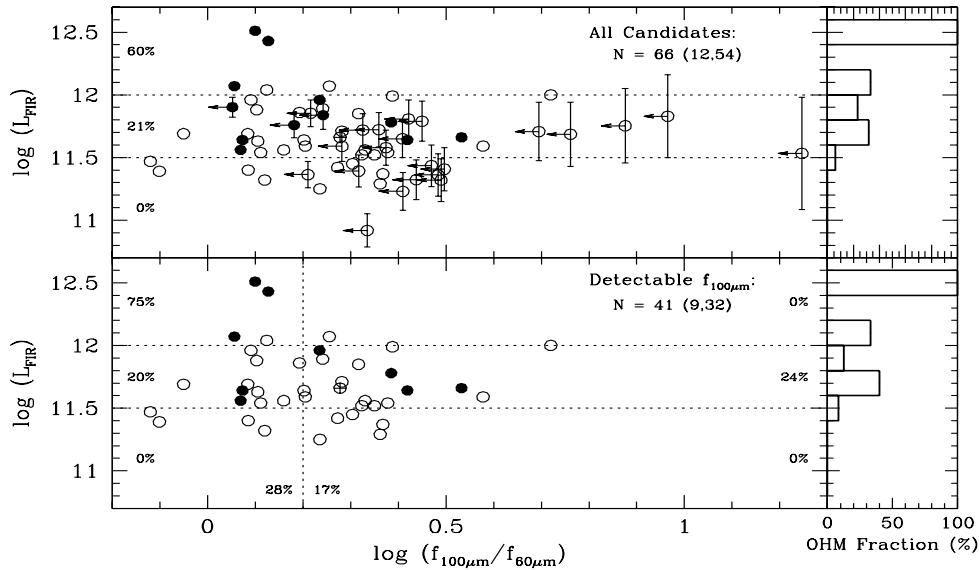}{3in}{0}{70}{60}{-220}{-190}
\caption{Observed OH Megamaser Candidates.  The two left panels show $L_{FIR}$
versus FIR color for candidates observed to date, and the two right panels
show the OHM fraction as a function of $L_{FIR}$.  Filled circles mark OHMs, 
empty circles mark non-detections, and the crossed circle marks the OH
absorber.  Points with error bars are non-detections at 100 $\mu$m.  
Vertical error bars indicate the possible range of $L_{FIR}$, constrained
by $f_{60\mu m}$ and an upper limit on $f_{100\mu m}$.  Horizontal arrows
indicate upper limits on FIR color.  Inset percentages indicate the OHM
fraction for each sector delineated by the dashed lines.  The upper panels
plot all 65 candidates observed, plus one known OHM reobserved to check
the observing setup in April 1999.  The lower panels plot the 41 objects
with detected $f_{100\mu m}$.  The inset numbers follow the key:  
N $= $ Observed (OHMs, Non-Detections).  We use $H_\circ = 75$ km s$^{-1}$
Mpc$^{-1}$ and $q_\circ = 0$.}
\end{figure}

This survey will provide the {\it first} uniform, flux-limited sample of 
OHMs, finally making it possible to empirically explore the physics of
OHM phenomena in a statistically meaningful manner, and to evaluate 
theoretical models.  The uniform sample will also offer the first 
reliable measure of the incidence of OHMs in ULIRGs as a function of 
host properties, especially $L_{FIR}$.  
Once the OHM fraction in ULIRGs is quantified at low redshifts, 
one can perform OHM surveys at higher redshifts to measure the luminosity 
function of ULIRGs --- and hence the merger rate of galaxies --- at arbitrary 
redshifts.  The current survey will be able to measure the galaxy merger
rate up to $z=0.25$.  

Our preliminary result demonstrates the high OHM detection rate achievable
with the upgraded Arecibo telescope in short integration times.  
The strong dependence of OHM fraction in ULIRGs on $L_{FIR}$ and
the selection of the most FIR-luminous LIRGs (via the criterion 
$z>0.1$) produce a high detection rate compared to previous OHM 
surveys (e.g., Staveley-Smith et al. 1992; Baan, Haschick, \& Christian
1992).  By extrapolation, we predict that {\it most} hyperluminous IR
galaxies host detectable OHMs, and that an OHM survey of this class
of LIRG would be highly successful.  The main barrier to this type of
survey is the paucity of detectors at the redshifted frequency of OH 
in the range $0.5 < z < 1.5$.  

ALMA will resolve the molecular cloud structures 
in galactic nuclei which produce OHMs and provide an unprecedented 
understanding of the dynamics, density field, and composition of these
regions.  Tying molecular studies of low redshift galactic nuclei to 
the corresponding observed OHM properties will allow us to extend that
local understanding to a physical study of OHMs at high redshift.  

\acknowledgements

The authors are very grateful to Will Saunders
for access to the PSCz catalog
and to the staff of NAIC for observing assistance and support.  
This research was supported by Space Science Institute archival grant 
8373 and made use of the NASA/IPAC Extragalactic Database (NED) 
which is operated by the Jet Propulsion Laboratory, California
Institute of Technology, under contract with the National Aeronautics 
and Space Administration.

\end{document}